# Interdisciplinary Expertise to Advance Equitable Explainable AI

**Authors**
Chloe R. Bennett*,[1,2], Heather Cole-Lewis*,[†, 3], Stephanie Farquhar[3], Naama Haamel[3], Boris Babenko[3], Oran Lang[3], Mat Fleck[3], Ilana Traynis[4], Charles Lau[4], Ivor Horn[‡,3], Courtney Lyles[‡,3,5,6]

[1] Work done at Google via Pro Unlimited, Folsom, CA, USA
[2] Northeastern University, Boston, MA
[3] Google, Mountain View, CA
[4] Work done at Google via Advanced Clinical, Deerfield, IL, USA
[5] University of California San Francisco, Department of Medicine, San Francisco, CA USA
[6] Center for Healthcare Policy and Research, University of California, Davis, Sacramento, CA

## Abstract

The field of artificial intelligence (AI) is rapidly influencing health and healthcare, but bias and poor performance persists for populations who face widespread structural oppression. Previous work has clearly outlined the need for more rigorous attention to data representativeness and model performance to advance equity and reduce bias. However, there is an opportunity to also improve the explainability of AI by leveraging best practices of social epidemiology and health equity to help us develop hypotheses for associations found. In this paper, we focus on explainable AI (XAI) and describe a framework for interdisciplinary expert panel review to discuss and critically assess AI model explanations from multiple perspectives and identify areas of bias and directions for future research. We emphasize the importance of the interdisciplinary expert panel to produce more accurate, equitable interpretations which are historically and contextually informed. Interdisciplinary panel discussions can help reduce bias, identify potential confounders, and identify opportunities for additional research where there are gaps in the literature. In turn, these insights can suggest opportunities for AI model improvement.

---
*Equal Contributions.
‡Equal Leadership.
† Corresponding author: hcolelewis@google.com

# Introduction

Artificial intelligence (AI) and machine learning (ML) are increasingly being used in healthcare and have changed how patient and population health is managed. Although much progress has been made in the AI and ML fields with regard to improving model performance for healthcare applications, many argue AI has the potential to perpetuate or even exacerbate health inequities (Liu et al., 2022; Rajkomar et al., 2018; WHO, 2021). In fact, many empirical studies have shown the power of AI to cause harm. Obermeyer et al. (Obermeyer et al., 2019) famously highlighted the racial bias present in an algorithm commonly used in healthcare. Bias is a major concern in healthcare AI, and many social scientists argue it arises due to the lack of historical and contextual data on structural factors, such as racism or sexism, as well as biases ingrained in both the medical field and within data collection and organization broadly (Cao et al., 2022; Inker et al., 2021; Robinson et al., 2020), especially in electronic health records and other administrative databases that were not designed to produce a representative sample of a population (Getzen et al., 2023; Himmelstein et al., 2022; Rozier et al., 2022; Sun et al., 2022). This concern is especially salient as the field moves toward providing information that can influence human behavior more directly, for example, by providing clinical decision support or empowering people to better understand and manage their conditions via health informatics and predictive analytics.

The use of techniques for the interpretation and explanation of AI models (broadly defined as explainable AI, or XAI here) has been proposed as an approach to promoting the equitable use of AI in healthcare (Yoon et al., 2022). Several methods have been proposed, including input perturbation (Zeiler & Fergus, 2014) and the analysis of neuron activations (Erhan et al., 2009) or gradient information (Selvaraju et al., 2017). However, such methods generally do not enable reliable hypothesis generation about the social, structural, and physiological processes underlying the observed associations between model inputs and the target of prediction (for example, historical Jim Crow laws are associated with disproportionately high rates of Black infant deaths in the U.S.;Krieger et al., 2013); nor do they necessarily faithfully explain the mechanism of the computation of an AI model, (Adebayo et al., 2018; Ghassemi et al., 2021; Lipton, 2018; Rudin, 2019).

We argue that a focus on the social and structural causes of health inequities can be useful in thinking about how to ask questions about the systems and structures that may have produced an AI model's results, and to more fully understand the model output, improving the practice of XAI. In this work, we introduce a framework for interdisciplinary expert panel (IDEP) review of XAI analyses.This framework allows for an equity-focused, critical assessment of the interpretation of the results of XAI analyses, identification of potential biases, including inaccurate causal language or logic used in the interpretation of findings, and can suggest directions for future research. We start by providing context on the need for an IDEP and explain how our approach to dissecting bias was informed by the fields of social epidemiology and health equity.

## Background

Robinson et al. (Robinson et al., 2020) argue that predictive modeling in health and healthcare will have causal impacts: "*...Model results will be used for decision-making about the allocation of health care, access to social welfare and disability systems, and acceptable limits of medical exposures for vulnerable populations.*" Because of this, many epidemiologists have argued that methods that use logic from the field of causal inference are necessary when using the results from predictive models. This becomes particularly important when results from a model are interpreted in a way that may imply causality. It is especially important to do this in health because hypotheses that explain health phenomena often use the language or metaphor of biology and can reinforce the 'naturalness' of unjust social phenomena (to take one example, in the 19th-century clinicians and scientists created and published a hypothesis in reputable medical journals that a disease called "drapetomania" caused enslaved people to run away and that whipping was both primary and secondary prevention for drapetomania (Bassett & Graves, 2018)).

Building a similar practice in AI can start to address some of the epistemological challenges of AI's so-called "black box" (Magrabi et al., 2019; Wadden, 2022). Within healthcare, the use and interpretation of AI models can lead to a confusion of "shortcut features," which are spuriously associated with a disease, with ones indicative of true disease pathology (Brown et al., 2023; Geirhos et al., 2020). Causal inference can be used as a central guiding framework for addressing issues related to bias through statistical and domain expertise to uniquely identify the effects of interest using observational data (Petersen & van der Laan, 2014). In the context of AI, causal inference can provide a formal framework to facilitate statistical estimation aligned with the underlying concept the analysis is targeting. When interpreting results from AI models, various causal inference frameworks (e.g. the potential outcomes framework; (Rubin, 2005)), can offer a set of logic that is useful for identifying ways an association may be biased.

***A Brief Overview of Common Biases and Their Relevance to Healthcare AI Development***

Electronic health record datasets used for AI model development can suffer from information bias when variables in the dataset are misclassified, mismeasured, or missing systematically. Misclassification can happen, for example, when a person is categorized as a racial category with which they do not identify. Measurement error may occur from human errors (e.g. incorrect data entry) and systematic differences across health systems (e.g. patients with lower socioeconomic status may frequent a clinic that uses different data and clinical reasoning practices (Gianfrancesco et al., 2018)). People participate in the healthcare system in different ways, which can affect the data available; these behavioral differences are often non-random. For example, patients who cannot access or navigate an online health system are less likely to have data on patient-reported outcomes (Gianfrancesco et al., 2018).

Selection bias can arise when there are systematic unobserved differences between the characteristics of the study sample and the study's target population (Bareinboim & Pearl, 2012). For example, a dataset taken from a hospital in the southern U.S. might contain a higher prevalence of diabetes and therefore may produce predictions of various comorbidities at different rates than that observed among a population with a lower prevalence of diabetes. Relatedly, AI models are also subject to issues of generalizability which often occur when a dataset used to train a model is not representative of the population of interest, as is often the case with datasets not created for research purposes; for example, electronic health records (Finlayson et al., 2021; Sperrin et al., 2022). As many predictive models may be highly accurate given one dataset, they may not perform well in the real world and across groups because of lack of generalizability due to the limits of the specific sample in the dataset(s) used or design choices made during model development (D'Amour et al., n.d.; Hooker, 2021).

Within any modeling approach that may be used to infer causality there are additional critical equity-specific methodological considerations, especially for variables representing social constructs. It is clear that XAI may report associations that could indicate racial, ethnic, gender, sexuality, ableist, nativist, or class bias in the data, and therefore attention to these considerations are crucial for the development of equitable AI as to not perpetuate biased notions of causality, such as the biased belief that Black people have thicker skin (Hoffmann, 2019). Such equity-specific modeling considerations for XAI include 1) explicit clarity and reporting of the definitions of variables for race/ethnicity, gender, identity and expression, sexual orientation, ability, and socioeconomic status, including documented investigation into the validity of these variable definitions, 2) considerations for complex interactions between variables and assessment for non-random patterns of missingness, and 3) clarity around gaps in data collection potentially driving XAI findings. These factors could apply to inputs to XAI methods, target predictions of the methods, or both.

To illustrate these considerations, we use race for many examples, which is used often in algorithms despite vague and inconsistent definitions of racial categories (Hanna et al., 2020) and despite the fact that there is no biological basis for racial categories, although exposure to racism has biological harms (Bailey et al., 2017, 2021). In data pre-processing, subgroups with a lower sample size are often aggregated into one vaguely defined group (e.g., non-White) which can obscure the relationships that a variable like race may have with other variables and result in subgroups being removed from analysis (Lett & La Cava, 2023). Complex interactions are also common, especially as variables are interconnected and influenced from the individual through community and societal levels. The disproportionately high rates of maternal mortality among Black women are further compounded by income level, with those in the lowest income level experiencing the most severe maternal outcomes among women who identify as Black (Singh & Lee, 2021; Vilda et al., 2019). These racial and income disparities are the result of complex structural discrimination underpinned by historical racism (Crear-Perry et al., 2021). Further, social categorization may often be inaccurate and non-generalizable. For example, misclassification of race within electronic health records has been documented to affect people who identify as Black more so than people who identify as White (Klinger et al., 2015), and missing data for race is less common for people who identify as White, therefore adding another structural advantage to whiteness (Cowger et al., 2020; Labgold et al., 2021). This data misclassification and non-random patterns of missingness can affect the strength and direction of observed associations.

Several methods have been proposed for bias detection in AI, most of which are reliant on statistical techniques for the identification of bias or provide domain-agnostic checklists with prompts designed to assist developers in incorporating ethical and responsible best practices (Cary et al., 2023; Richardson & Gilbert, 2021). Some have proposed the need for domain-specific checklists or guidelines which are useful for the end-user (Atkins et al., 2021), while others have argued for the importance of participatory design and community engagement (Birhane et al., 2022; Y. Chen et al., 2023; Katell et al., 2020; Prabhakaran & Martin, 2020; Young et al., 2019). Algorithmic impact assessments (Reisman et al., 2018) and algorithmic auditing (Liu et al., 2022; Raji et al., 2020) have also been proposed. While these equity-focused methods have seen success in the healthcare space, such as in cardiovascular health among an African-American community (Harmon et al., 2022), they have not gained widespread popularity due to many structural factors such as lack of standards and legal regulations (AI Now Institute, 2018; Hoffmann, 2019).

# Methods

The IDEP framework for interdisciplinary expert panel review of XAI analyses was developed by applying existing frameworks from health equity research that build on social epidemiology as a core discipline to XAI. The IDEP framework was then applied to a concrete case study to solidify an interdisciplinary approach to identification of potential biases and directions for future research.

**Interdisciplinary Expert Panel (IDEP) Review Framework**

Our framework for IDEP is presented in Figure 1. The framework was informed by the World Health Organization's Social Determinants of Health Model (WHO, 2010), Bronfenbrenner's Ecological Systems Theory (Bronfenbrenner, 1992), and Krieger's Ecosocial Theory (Krieger, 2011). These models and theories explain the social, political, economic, intergenerational, life course, and environmental factors that shape health, positing that health is not simply biologically determined, although biology is affected by these factors. Using these models and theories, we designed the IDEP framework to intentionally place societal and political, communal, organizational, and individual contexts at the forefront of expert discussion. Prompts to aid panel members in identifying relevant context-specific evidence and rationale are then used to examine and discuss model findings and whether they are due to change, bias, or confounding. Table 1 provides examples of social and structural factors relevant to health across various socio-ecological levels that may be useful for panel member consideration.

To demonstrate the IDEP framework for the interpretation of XAI findings, we use a case study of an XAI method that applies a generative model to explain predictions made from medical imaging data (StylEx). A full description of the StylEx method can be found elsewhere (Lang et al., 2024). The StylEx method consists of 4 fundamental steps: 1) the AI researchers train a classifier on images, 2) the AI researchers train a "StylEx" model – a StyleGAN-based image generator guided by the classifier, 3) the StylEx model automatically extracts the top attributes which affect the classifier, and 4) the AI researchers generate a visualization of these attributes for expert evaluation, discussion, and consensus. For the purposes of this paper, we focus discussion on step four, in which clinicians identify the phenomena from the visualization that they believe is producing the prediction. This step is a human hypothesis interpretation of how the model works, which we argue must be tested by an interdisciplinary expert panel using causal reasoning to consider forms of "bias," e.g., social or structural factors or methodological decisions that might have contributed to the prediction.

The StylEx work studied several AI models for predicting various disease states or physiology measurements from medical images: fundus photographs, external eye photographs, and chest radiography. StylEx was applied to generate visual attributes (in the form of animated images) that most influenced the predictive AI models. For example, specific vascular changes were observed to increase the predicted probability for male biological sex in fundus photographs and increased eyelid margin pallor increased the predicted probability for glycated hemoglobin (a marker of blood sugar levels). In some cases, the model produced results that may be representative of social or structural factors' influence. For example, the model's observed association between eyelid makeup / eyeliner and low hemoglobin may be interpreted as the model detecting the association between menstruation and anemia, with the underlying assumption being that people who menstruate are more likely to wear eye makeup. However, one must consider various confounding factors that may have produced this association among the dataset used for this model's prediction. Such ponderance leads to questions about the representativeness of the data and whether this association would be observed in a dataset comprised of only female biological sex or only those with darker skin tones.

This paper introduces the IDEP review framework leveraged in the StylEx study, which aims to explain observations by assessing the plausibility of biological, clinical, or social pathways linking model predictors to outcomes. Using this as a case study, we examine findings from an XAI method that identifies image attributes linked to predictions of clinical/biological outcomes. The IDEP method has potential to uncover bias at multiple levels: structural bias (e.g., racism, sexism, classism) present in data collection and processing, scientific racism present within medicine/healthcare, and common statistical biases like selection bias and confounding.

## Results

### Case Study: IDEP Review of StylEx Findings in Healthcare
*Selection of Interdisciplinary Subject Matter Experts for Panel Participation*

The panel consisted of AI researchers and subject matter experts (N=9). The selection of subject matter experts was informed by expertise (i.e., academic training, practice-based knowledge, lived experience) in five criteria: 1) the type of dataset, 2) the setting(s) where the data were collected, 3) the domain(s) where the model findings are applicable (e.g., healthcare, research), 4) design research in public health, healthcare, or medical devices, and 5) causal inference. In this case study using the StylEx method (Lang et al., 2024), because the XAI method uses data collected in a healthcare setting via radiology and ophthalmology imaging devices, and the model findings could be used to inform medical practice, we identified potential panel members with experience and expertise relevant to the subjects related to radiology, ophthalmology, human factors engineering, and social sciences. Our panel thus comprised two clinicians, one user experience researcher, and two social scientists with expertise in epidemiology, behavior science, and health research, in addition to the StylEx model developers ("AI researchers" hereinafter). A summary of each panelist's subject matter expertise, including the AI researchers, can be found in Table 1. Clinicians comprised a radiologist with a subspecialization in thoracic conditions and a comprehensive ophthalmologist, both with experience in clinic and hospital settings. The user experience researcher, whom we refer to as the socio-technical scientist, has conducted many studies in hospital and clinic settings with patients, physicians, and tech assistants to aid in the development of medical devices, such as imaging devices, often used in healthcare settings. One social scientist has experience in behavior science and epidemiology research. The second social scientist has experience in epidemiology and health services research. Both social scientists have experience in health equity research focusing on the social and structural determinants of health. All social and sociotechnical scientists have an advanced understanding of causal inference and have experience with quantitative and qualitative health research. Additional expertise considered included social work, infectious disease epidemiology, genetic epidemiology, and industrial engineering. However such experts were not included in the panel as we felt such specialized expertise was not necessary for this specific AI model and our panel members all met the five criteria presented above. Additionally, panelists exhibited a diverse array of racial, ethnic, and gender identities, as well as varied lived experiences.

*IDEP Process Overview*

Figure 1 summarizes the four phases of the IDEP review of XAI findings. In the first phase, clinicians review findings from the XAI method with the AI researchers to describe the visual attribute(s) observed in the XAI finding based on anatomical and clinical knowledge. For each XAI finding (i.e., an image attribute that is being identified by the XAI method as having an association with the model predictions), the clinicians work with the AI researchers to establish the visual signal assumed to be captured by the image attribute in plain language (i.e., describe the variation in the image attribute in a human understandable fashion such as, "eyelid margins become paler"), identify and discuss relevant pathophysiological links to the image attribute, and explore their congruence, based on existing medical knowledge and experience. In the second phase, initial explanations are presented to the social and socio-technical scientists for review and each person provides additional evidence based on experience, empirical research, and/or literature in support or conflict of the initial explanations. All panelists are encouraged to consider all plausible instances of bias, including confounding variables. Additionally, during phase two, the social and socio-technical scientists develop questions about the data validity, data collection environment, statistical methods, and historical/current contexts. In phase 3, the full expert panel, including the AI researchers, reviews the posed questions developed in phase 2 and generates hypotheses for why the signal is useful to the classifier. In the last phase, the subject matter experts and AI researchers together consider all evidence raised in phases 1-3 in support or conflict of initial explanations and come to a consensus on the best way to summarize the XAI findings.

*Interpretation of XAI Findings (Phases 1-3)*

Although the panel discussed all image attribute associations from the XAI method, specific focus was given to attributes that the panel identified as potentially being representative of bias. The panel placed high importance on discussing the potential for bias introduced via systemic or structural factors present in the dataset or confounding not accounted for in the model. For example, one XAI finding was a strong association between upper eyelid/eyeliner makeup and low hemoglobin. At first review, the clinicians provided insight into the known association between anemia and menstruation, suggesting a hypothesis that the model may be able to detect the likelihood of low hemoglobin (indicative of anemia) from the presence of eyeliner, considering that eyeliner use is assumed to differ across biological sexes. During the social and socio-technical science review, questions were raised regarding environmental factors associated with both eyeliner use and anemia (i.e., confounders), the representativeness of the dataset, and the generalizability of the associations. In this instance, the IDEP discussed possible explanations beyond biological ones, which could explain the association between the presence of eyeliner and low hemoglobin levels. Panel members discussed whether the association would still be present if the model were run on datasets that were filtered by a rough proxy for eye makeup usage (i.e., biological sex) or if the model had a greater distribution of images with varying levels of contrast (e.g., a dataset including varying skin tones using dark eyeliner or dataset including people of similar skin tones using metallic or lighter eyeliner colors). Table 1 lists examples of social and structural determinants the panel considered in their discussions.

*Categorizing of XAI Findings (Phase 4)*

Once the discussion of the XAI associations concluded, the panel came to a consensus on how to categorize each finding. Associations were categorized as either *a) known association, b) known in the clinical literature; could warrant further statistical examination, c) novel association; warrants further investigation, or d) strong likelihood of confounding or other bias present; further investigation highly recommended*. Additionally, the full panel identified a list of research and validation hypotheses that can be tested either by comparing them to known phenomena from the literature, or by designing further studies which can prove or disprove them. The first category, *known associations*, was composed of associations agreed upon within the clinical literature that are unlikely to be confounded by social or structural factors. For example, one association was noted between decreased conjunctival vessel prominence (small blood vessels on the whites of the eye) and low hemoglobin. This decrease in vessel prominence is a well-known sign of anemia attributed to a reduction of circulating oxyhemoglobin in tissue vasculature and therefore unlikely to be affected by social or structural factors. The second category, *known association which warrants further statistical examination*, consisted of associations previously observed in the medical literature but which the panel felt may be artifacts of social or structural factors rather than biological processes. One such example was the association between retinal vein dilation and being a smoker. The clinical literature has found such associations previously, but the evidence does not rule out the effects that cardiovascular disease, diabetes, or other conditions often associated with smoking may have on eye health. Thus, in our categorization, the panel recommended that further statistical examination be conducted to disentangle the complex relationship between smoking, cardiovascular disease, and diabetes and better isolate a less biased or confounded association with smoking. In the *novel association* category, panel members agreed such associations warranted further exploration to better understand why the model found such associations. For example, the model found an association between the conspicuity of skeletal structures (e.g., vertebrae, ribs, scapulae, and humeri) relative to background soft tissue and Black race (self-reported by patients using a multi-select question; in analyses data were binarized to be Black and non-Black to match prior literature race prediction setups). An assessment not historically or contextually informed might conclude an association between bone density and Black race. Given the historical scientific racism still present in medical literature and research today (Fausto-Sterling, 2008; Smedley & Smedley, 2005), as well as encoded in contemporary clinical practice through longstanding clinical algorithms (Neal & Morse, 2021; Vyas et al., 2020), and the widely accepted understanding that race is socially, rather than biologically, constructed (AMA, 2020; Krieger, 2002; Romualdi et al., 2002), this finding was critically interrogated by the panel. In the last category, the panel labeled associations appearing to be highly confounded as, *strong likelihood of confounding or other bias present; further investigation highly recommended*. For example, the model found a negative association between increased upper lung volume and Black race. One panel clinician noted that increased lung volumes are associated with COPD, which is more prevalent in White populations than in Black populations. However, upon further discussion of social and structural determinants, the panel concluded that this finding was likely confounded by the racial disparities in diagnostic/screening rates for COPD (e.g., screening and diagnosis are lower among Black populations; (Mamary et al., n.d.). Additionally, another XAI finding was that apparent image overexposure increases the probability of a chest X-ray being classified as "abnormal", which can be associated with the X-ray projection used to acquire the chest radiograph (e.g. X-ray tube anterior to the patient with the image detector posterior to the patient, or vice versa). The projection in which a chest radiograph is obtained can be associated with factors such as patient mobility, acuity, inpatient vs. outpatient status, and other social or structural factors which are not noted in the dataset. In cases like this, the panel concluded that potential avenues for further investigation might include re-training the model using a different dataset or subsets of the data, assessing the effects of confounding variables such as by presenting results stratified by race and age combined, or augmenting the dataset with data from public health datasets that include factors at the individual, societal, political, and geographical levels.

## Discussion

Clinical research and public health practice have long grappled with the ability to use advanced methodological approaches to 1) understand risk factors within the population, 2) intervene to reach individuals or communities most effectively, and 3) evaluate the performance of clinical interventions or programs, overall and within subgroups. There is a long history of spurious associations and harmful interpretations of analyses, especially related to social constructs such as race/ethnicity and gender. AI for health and healthcare should be able to differentiate associations and signals from causal relationships, prioritizing thorough and rigorous validation in diverse populations and the environments where it's deployed in order to have maximum impact for real-world practice or decision making (Ghassemi et al., 2021; Sendak et al., 2020). AI differs from the typical historical process of generating clinical and public health evidence, given that AI can learn and adapt to patterns directly from minimally-curated observational data and is uniquely positioned to rapidly integrate new data to improve model performance. However, the same core issues of bias data and interpretations remain; this is especially concerning when considering the growing use of AI for clinical decision support and diagnostics. In this paper, we present a method for interdisciplinary expert review of XAI findings which help to evaluate the plausibility of biological, clinical, or social pathways contributing to observed associations between the model predictor and outcomes. We use a case study of findings from an XAI method which uses medical images to identify specific image attributes associated with clinical/biological outcomes. The expert panel review concluded that several of the associations the XAI method found may be due to dataset demographics, human interactions with technology, or the result of social and structural determinants or bias. The expert panel also concluded that some associations might reveal novel phenomena and suggested research to support such discoveries.

      Our proposed framework provides a guide for how to critically assess XAI findings for bias, generalizability, and equitability by leveraging expert knowledge, including from lived experience, to guide the review of XAI findings. This framework enabled the identification of common biases present in electronic health records. For example, measurement error and misclassification, arose in the discussion of race in the model being coded as binary White or Black. Selection bias was discussed with regard to the subset of people who would receive retinal imaging services (e.g., a sample with a higher prevalence of diabetes and access to healthcare or a sample with a higher prevalence of infectious lung diseases). Lastly, AI researchers' decisions for inclusion/exclusion criteria used to identify a subset of data to train a model may introduce bias. For example, our panel discussed the fact that a patient with limited physical mobility (e.g., cannot stand, cannot lift their arms) may require a diagnostic imaging protocol that is different from a patient with full mobility. There are many more instances where bias can occur and within the scope of this paper, we are unable to cover each in-depth. Thus, we refer interested readers to others who have written more extensively about how to incorporate causal thinking into model development from model conception to interpretation (Petersen & van der Laan, 2014).

This work provides a framework that marries two methods relevant to XAI, deductive reasoning and inductive learning (Aggarwal, 2021), allowing us to construct hypotheses from empirical data that are informed by expertise in medicine, healthcare, epidemiology, human factors engineering, and data science. Using these methods in tandem addresses the limitations of both by using extensive, evidence-based expertise in how structural discrimination (e.g. racism, sexism, classism, ableism) functions within society and the US healthcare system to guide our understanding of the patterns of observations data produce. The framework presented was informed by the World Health Organization's Social Determinants of Health Model (WHO, 2010), Ecological Systems Theory (Bronfenbrenner, 1992), and Ecosocial Theory (Krieger, 2011), and allows for examination of the social, political, economic, and environmental forces which shape health that an XAI method cannot uncover alone. A framework such as the IDEP framework may be particularly useful for AI models in healthcare, including generative AI, which pose the risk of providing inaccurate diagnoses or medical information and perpetuating bias and which could be more prone to being misinterpreted as describing causality (R. J. Chen et al., 2021; Ktena et al., 2023; Luccioni et al., 2023; Zhang et al., 2020).

The interdisciplinary expert panel review framework can be useful in tandem with other proposed methods for bias detection via transparency artifacts, such as datasheets, datacards, model cards, and healthsheets (Gebru et al., 2021; Mitchell et al., 2019; Pushkarna et al., 2022; Rostamzadeh et al., 2022) which can be presented to panel members and used to guide discussions and assess whether associations are the result of bias present in the dataset. This method can easily incorporate aspects of community-based participatory research by including community members or patient advocates as panel members whose subject matter expertise lies in their lived experience navigating their health and the healthcare system or supporting others doing the same. Inclusion of community members or patient advocates is in line with the 5 criteria for expertise to be included in the IDEP, specifically criterion 3– expertise in the domain(s) where the XAI findings are applicable. This process of applying interdisciplinary panel reviews to XAI findings represents an additional method for future policy considerations to address bias in health AI.

*Limitations*

This method has limitations. Gathering a panel of interdisciplinary experts can be a difficult task and not all AI practitioners will have access to individuals with social and socio-technical expertise. We encourage AI practitioners to leverage their larger professional networks and seek collaboration from those in academia, public health, or healthcare and engage in community engagement to support the involvement of community members and patient advocates. In addition to focusing on diversity as it relates to subject matter expertise and professional experience, we stress the importance of assembling a panel of experts with diverse social identities, perspectives, and lived experiences. Additionally, it is important to consider that social identities are multi-faceted and cannot be summarized simply by, for example, racial or gender identity. In this study, the expert panel, by chance, included panelists with diverse perspectives and ability to speak to a myriad of lived experiences in addition to professional expertise; however there was no deliberate effort to recruit for diversity along the dimensions of social identity, perspectives, and lived experiences. Future work may consider explicit recruitment based on factors that are relevant to the subject matter. For instance, in this study, given the application of the StylEx method in healthcare and acknowledging the persistence of racism in the healthcare system, it would have been prudent to recruit specifically for panelists who identify as Black with lived experiences related to racism or some similar intersectional identity. Examination of the societal context of the healthcare system may have uncovered other insightful social identities, perspectives, and lived experiences to have been represented by experts on the panel. A list of factors for consideration in US societal contexts include: skin color, race, ethnicity, gender, sexual orientation, class, ability/disability, religion, age, language, education, documentation, and insurance status.

      Panel reviews are typically non-exhaustive and are subject to the biases present in the experience and training of each individual panelist. However, a proper panel composed of experts with unique or maximally non-overlapping professional and lived experiences can ensure panelists are able to produce a rich discussion and interrogation of XAI findings. For example, an epidemiologist might provide context for how the prevalence of type 2 diabetes complications is higher among people using public insurance (e.g., Medicaid, Medicare). At the same time, a qualitative health researcher with a background in social work may leverage their understanding of the complexities of medication coverage/approval for patients using public insurance. Our method also relies on the ability of the AI researchers to explain how the model works and to be knowledgeable about the dataset to answer questions about the structure of the data. Again, we suggest the use of transparency artifacts to address questions about dataset structure. As with any method reliant on observational data, it is impossible to identify and evaluate all forms of bias, thus the possibility of bias remains and each hypothesis generated would require validation through additional quantitative research to confirm the outcomes. Additionally, validating generated hypotheses may prove difficult given the dearth of data with variables at multiple levels (e.g. political and geographic). Such datasets would be

useful in determining if IDEP results are uncovering additional hypotheses that are accurate or if they primarily identify confounding variables. Lastly, this method also relies on a high level of mutual respect between panelists and strong facilitation skills to defuse defensiveness or relying on privilege and power in ways that do not serve the scientific agenda.

## Conclusion

AI is rapidly influencing healthcare, and the biases it can produce and perpetuate are well-documented. We emphasize the importance of deconstructing the associations that XAI methods find and the interpretation of those findings. To operationalize this, we present a method for an interdisciplinary expert panel review that discusses an XAI method output, critically assesses the XAI findings, and identifies areas of bias. This method may yield more accurate, equitable models, which are historically and contextually informed, by reducing bias, identifying potential confounding, suggesting areas for model improvement, and identifying opportunities for additional research where there are gaps in the literature.

The promise of AI that promotes health equity is likely not achievable purely through big data and AI methods. Collaboration with social and socio-technical scientists, including qualitative researchers, via expert panel review can promote AI models which translate data into useful, equitable healthcare tools. With expert panel reviews, we can potentially uncover bias at multiple levels: structural bias (e.g., racism, sexism, classism) present in data collection and organization/processing, scientific racism still functioning within the practice of medicine, and common statistical biases like selection bias and confounding. With this method, panelists are situated in a setting that facilitates discussion and knowledge sharing, allowing for the surfacing of potential biases. While there is no silver bullet that allows us to produce AI rid of bias, we suggest the use of this method, in tandem with other previously suggested methods, to help limit and address the bias and harms perpetuated by healthcare AI by actively identifying and fixing issues, enabling fundamental change.

## Acknowledgements

The authors would like to thank Stephen Pfohl, Jessica Schrouff, Sami Lachgar, Lauren Winer, and Dale Webster for their feedback and support of this work. This study was funded by Google LLC.

# Tables

**Table 1. Examples of Social and Structural Determinants Considerations to Guide Experts during IDEP Review Process**

| Socio-ecological Levels | Social and Structural Determinant Considerations (non-exhaustive list) |
|---|---|
| Society and policy | <ul><li>Healthcare payment structures/reimbursement</li><li>Recommendations for screening/treatment</li><li>Changes in regulation of medical devices</li><li>Changes in medical device design trends</li><li>Economic and social policies</li><li>Structural discrimination</li></ul> |
| Community | <ul><li>Access to care</li><li>Insurance coverage</li><li>Trust in healthcare</li><li>Acceptance and trust in new technology</li><li>Access to transportation</li><li>Community safety</li><li>Health promotion</li></ul> |
| Organizational | <ul><li>Average appointment length</li><li>Tech innovation / implementation</li><li>Willingness to change</li><li>Physical space where technology is used</li><li>Quality of care</li><li>Disparities in screening and treatment</li></ul> |
| Individual | <ul><li>Medication adherence</li><li>Biological / genetic factors</li><li>Language and literacy</li><li>Patient interactions with tech</li><li>Physician / staff interactions with tech</li><li>Exposure to discrimination</li><li>Food security</li><li>Affordability of care</li></ul> |

**Table 2. Case-study Subject Matter Experts Details**

| Expert | Expertise |
|---|---|
| A[1] | Machine learning, Computer Vision |
| B[1] | Machine learning, Computer Vision |
| C[1] | Ophthalmology<br>Clinical research<br>Healthcare clinic environments and processes |
| D[1] | Biomedical research across medical specialties and modalities (molecular, records, signals, image, text, etc)<br>Machine learning (prediction models and XAI)<br>Healthcare (first responder) |
| E[2] | Radiology<br>Healthcare clinic environments and processes |
| F[2] | Ophthalmology<br>Healthcare clinic environments and processes |
| G[1] | Human-computer interactions<br>User experience design<br>Healthcare clinic environments and processes<br>Imaging technology<br>Psychology<br>Neuroscience<br>Causal inference |
| H[1] | Behavior science<br>Health equity<br>Health services<br>Epidemiology<br>Biomedical informatics<br>Causal inference |
| I[3] | User experience design<br>Health equity<br>Health services<br>Epidemiology<br>Causal inference |

1: Google, Mountain View, CA: No funding to declare.

2: Work done at Google via Advanced Clinical, Deerfield, IL, USA

3: Work done at Google via Pro Unlimited, Folsom, CA, USA; Northeastern University, Boston, MA; No funding to declare

## Figures

**Figure 1. Process Flowchart for Interdisciplinary Expert Panel (IDEP) Review Framework**

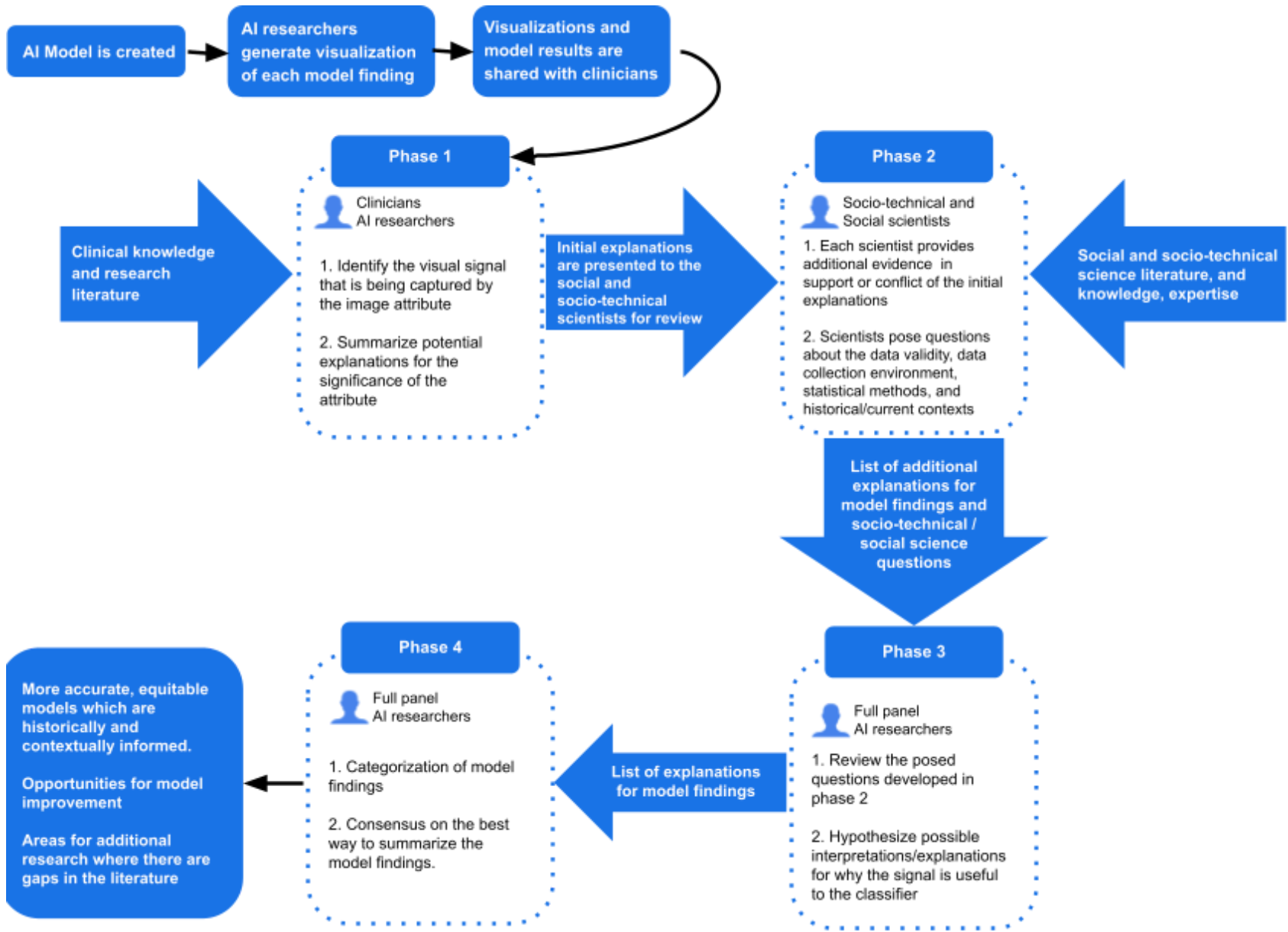